# The Virial Theorem and the Ground State Problem in Polaron Theory

N. I. Kashirina[a], V. D. Lakhno[b], and A. V. Tulub[c]

[a] Institute of Semiconductor Physics, National Academy of Sciences of Ukraine, Kiev, 03028 Ukraine
e-mail: n_kashirina@mail.ru
[b] Institute of Mathematical Problems of Biology, Russian Academy of Sciences, Pushchino, Moscow oblast, 142292 Russia
e-mail: lak@impb.psn.ru
[c] St. Petersburg State University, Universitetskaya nab. 7/9, St. Petersburg, 199034 Russia
e-mail: tulub@NK7099.Spb.edu

**Abstract**—The virial theorem for the translation-invariant theory of a polaron [3] is discussed. It is shown that, in [3], Tulub made a nonoptimal choice of variational parameters in the strong-coupling limit, which led to the violation of the virial relations. The introduction of an additional variational parameter to the test function reduces the polaron energy and makes it possible to satisfy the relations of the virial theorem for a strong-coupling polaron (the Pekar 1 : 2 : 3 : 4 theorem).

**DOI:** 10.1134/S1063776112030065

It is well known that one can often establish some general relations between the mean values of the kinetic, potential, and interaction energies for classical and quantum systems, which are known as the virial theorem. The virial theorem holds for both exact and approximate wave functions, provided that the functions are obtained by the variational method.

In particular, these general relations can be obtained for a polaron on the basis of the Frölich Hamiltonian:

$$H_p = -\frac{\hbar^2 \Delta_r}{2m} + \sum_k (V_k e^{i\mathbf{k}\cdot\mathbf{r}} a_k + \text{H.c.}) + \sum_k \hbar\omega a_k^\dagger a_k, \quad (1)$$

where $m$ is the electron effective mass, $\mathbf{r}$ is the electron coordinate, $a_k^\dagger$ and $a_k$ are the creation and annihilation operators of phonons with energy $\hbar\omega$,

$$V_k = \left(\frac{e^2 2\pi\hbar\omega}{k^2 \tilde{\varepsilon} V}\right)^{1/2}, \quad \tilde{\varepsilon}^{-1} = \varepsilon_\infty^{-1} - \varepsilon_0^{-1}, \quad (2)$$

$e$ is the electron charge, $\varepsilon_\infty$ and $\varepsilon_0$ are the high-frequency and static dielectric constants, and $V$ is the volume of the system.

The virial relations for the polaron problem for an arbitrary value of the electron–phonon interaction constant are given by [1]

$$T_p = -F_p, \quad E_{\text{el}} = 3F_p, \quad E_{\text{int}} = 4F_p, \quad (3)$$

where

$$F_p = T_p + E_{\text{int}}/2, \quad E_{\text{el}} = T_p + E_{\text{int}},$$

$$T_p = \langle \Psi_p | -\frac{\hbar^2 \Delta_r}{2m} | \Psi_p \rangle,$$

$$E_{\text{int}} = \langle \Psi_p | \sum_k (V_k e^{i\mathbf{k}\cdot\mathbf{r}} a_k + \text{H.c.}) | \Psi_p \rangle,$$

and $\Psi_p$ is the wave function of the ground state of a polaron. In the strong-coupling limit, one more relation is added to the virial relations (3):

$$E_{\text{ph}} = -2F_p, \quad (4)$$

where

$$E_{\text{ph}} = \langle \Psi_p | \sum_k \hbar\omega a_k^\dagger a_k | \Psi_p \rangle.$$

In this limit, $F_p$ coincides with the total energy of the ground state of the self-consistent state of a polaron (the thermal ionization energy) $E_p = \langle \Psi_p | H_p | \Psi_p \rangle$ and the virial relations obtained in [1] correspond to the well-known Pekar 1 : 2 : 3 : 4 theorem for a strong-coupling polaron [2].

In the translation-invariant theory [3], the field operators are subject to the translation transformation:

$$a_k \longrightarrow a_k + f_k,$$

here the function $f_k$ describes the classical field component (polaron well). In [3], based on the Hamiltonian (1), Tulub obtained the following expression for the ground-state energy $E_p$:

$$E_p = \Delta E + 2\sum_k V_k f_k + \sum_k f_k^2 \quad (5)$$

(as in [3], we put here $\hbar = \omega = 1$).

The quantities appearing in (5) have the following meaning:

$$T_p = \Delta E, \quad E_{int} = 2\sum_k V_k f_k, \quad E_{ph} = \sum_k f_k^2,$$

and should satisfy the virial relations (3) and (4). The following expression was obtained in [3] for $E_{int}$ with the use of the test function $f_k^T = -V_k \exp(-k^2/2a^2)$, where $a$ is a variational parameter:

$$E_{int} = 2\sum_k V_k f_k = -\frac{2}{\sqrt{\pi}} g^2 a, \quad (6)$$

where

$$g^2 = \alpha = \frac{e^2}{\hbar \varepsilon}\sqrt{\frac{m}{2\hbar\omega}}$$

and $\alpha$ is the Frölich electron–phonon coupling constant. Accordingly, the expression

$$E_{ph} = \frac{1}{\sqrt{2\pi}} g^2 a \quad (7)$$

was obtained for $E_{ph}$.

It follows from (6) and (7) that

$$E_{int}/E_{ph} = -2\sqrt{2}. \quad (8)$$

This expression contradicts the virial theorem (formula (4)).

In [3], the following value was obtained for the ground-state energy for a single variable parameter:

$$E_0 = -0.105 g^4, \quad (9)$$

which corresponds to a higher energy of the polaron compared with that obtained by Miyake [4],

$$E_0 = -0.10851128 g^4. \quad (10)$$

It is of interest to find a minimum of the total energy within the translation-invariant theory in the class of functions satisfying the virial relations. Within this problem, we have found that the choice of the test function $f_k^T$ made in [3] is not optimal. Functional (5) attains its minimum for a function of the type $f_k^0 = N f_k^T$ in which two variational parameters are specified. The variational parameter $N$ turned out to be equal to $N = \sqrt{2}$. For $f_k = f_k^0 = \sqrt{2} f_k^T$, the virial relations (3) and (4) are satisfied in a strong-coupling limit (the Pekar 1 : 2 : 3 : 4 theorem [2]), and, according to (5), the ground state energy takes the value

$$E_0 = -0.1257520 g^4, \quad (11)$$

which is much lower than the best value given by (10).

The value of $E_0$ given in (10) is currently a well-established value and is determined from the relation

$$\lim \frac{E_0}{\alpha^2} = \inf_{\Psi, \|\Psi\|=1} \left[ \frac{1}{2} \int d\mathbf{r} |\nabla \Psi(\mathbf{r})|^2 \right.$$
$$\left. - \frac{1}{\sqrt{2}} \int d\mathbf{r}_1 d\mathbf{r}_2 |\Psi(\mathbf{r}_1)|^2 |\mathbf{r}_1 - \mathbf{r}_2|^{-1} |\Psi(\mathbf{r}_2)|^2 \right], \quad (12)$$

where the functional on the right-hand side is the Pekar functional obtained in the strong-coupling limit. A rigorous proof of formula (12) was given in [5, 6]. Miyake's result has been reproduced in a large number of works (see [7, 8]) and does not raise any doubts.

In our view (the results of [3] have been checked once again), the only possible explanation to the arising contradiction is that one uses the wave functions belonging to different function classes in the translation-invariant theory in the strong-coupling approximation and in the strong-coupling theory based on the wave functions that minimize the functional (12). In the translation-invariant theory, the wave function of a polaron with zero total momentum is given by

$$\Psi_0^T = \exp\left(-i \sum_k \mathbf{k} a_k^\dagger a_k \mathbf{r}\right) \hat{\Phi}(\{a_k\})|0\rangle,$$
$$|\Psi_0^T|^2 = \text{const}, \quad (13)$$

where the explicit form of the functional $\hat{\Phi}$ is given in [3]. A transition to a localized description in the polaron problem (states in the theory with spontaneously violated translational symmetry) was also considered in [3] and yielded relation (9) for energy. The approximate wave function of the ground state defined by functional (12) belongs to the class of localized, normalizable, functions. At the same time, a rigorous substantiation of the delocalized character of the true wave function of a polaron in the ground state was given in [9].

In conclusion, note that the value (11) for the ground-state energy of a polaron suggests that one should re-evaluate the stability criterion with respect to the parameter $\eta_c = \varepsilon_\infty/\varepsilon_0$ for the bipolaron state $E_{bp} < -2E_p$ above which there are no bipolaron states. For the parameter $\eta_c$ obtained in [10] within the same quantum-field approach as the polaron energy given by (11), we obtain a value of $\eta_c = 0.179$ instead of $\eta_c = 0.2496$, which is calculated with the use of the polaron energy (10). Since in [10], when solving a bipolaron problem, an optimization has been carried out with respect to the additional variational parameter, the virial theorem for the bipolaron [11] is satisfied in [10] automatically.

ACKNOWLEDGMENTS

This work was supported by the Russian Foundation for Basic Research, project nos. 11-07-12054 and 10-07-00112.

*Translated by I. Nikitin*